\documentclass[12pt]{article}
\usepackage{amsmath}
\usepackage{graphicx}
\begin{document}
\baselineskip=18 pt
\begin{center}
{\large{\bf  Axially Symmetric Null dust Spacetime, Naked Singularity and Cosmic Time Machine}}
\end{center}

\vspace{.5cm}

\begin{center}
{\bf Faizuddin Ahmed}\footnote{faizuddinahmed15@gmail.com}\\ 
{\it Hindustani Kendriya Vidyalaya, Dinesh Ojha Road,}\\
{\it Guwahati-781005, India}\\
\end{center}

\vspace{.5cm}

\begin{abstract}

In this article, we present a gravitational collapse null dust solution of the Einstein field equations. The spacetime is regular everywhere except on the symmetry axis where it possesses a naked curvature singularity, and admits one parameter isometry group, a generator of axial symmetry along the cylinder which has closed orbits. The spacetime admits closed timelike curves (CTCs) which develop at some particular moment in a causally well-behaved manner and may represent a Cosmic Time Machine. The radial geodesics near to the singularity, and the gravitational Lensing (GL) will be discussed. The physical interpretation of this solution, based on the study of the equation of the geodesic deviation, will be presented. It was demonstrated that, this solution depends on the local gravitational field consisting of two components with amplitude $\Psi_2$, $\Psi_4$.

\end{abstract}

\vspace{0.2cm}


\vspace{.5cm}

\section{Introduction}

For algebraically special metrics, the Petrov classification is a way to characterize the spacetimes by the number of times principal null direction (PND) it admits. The various algebrically special Petrov types have some interesting physical interpretations in the context of gravitational radiation. In the Newmann-Penrose notations \cite{Newmann1,Newmann2} (null tetrad $(\bf {k, l, m ,{\bar m}})$, where $\bf {k, l}$ are real and $\bf {m ,\bar{m}}$ are complex conjugate of each other), if the tetrad vector $k^{\mu}$ is a principal null direction, then the algebrically special metrics automatically gives, $\Psi_0=0$. For the algebraically special metrics, the special cases are :
\vspace{0.1cm}
\begin{eqnarray}
&&\Psi_0=0=\Psi_1,\quad\Psi_2\neq 0\quad\quad\quad\quad\quad\,\,:\quad \mbox{type}\,\,II,\nonumber\\
&&\Psi_0=\Psi_1=\Psi_2=0,\quad\Psi_3\neq 0\quad\quad\,\,\,\,:\quad \mbox{type}\,\, III,\nonumber\\
&&\Psi_0=\Psi_1=\Psi_2=\Psi_3=0,\quad\Psi_4\neq 0\,:\quad \mbox{type}\,\,N,\nonumber\\
&&\Psi_0=\Psi_1=\Psi_3=\Psi_4=0,\quad\Psi_2\neq 0\,:\quad \mbox{type}\,\,D\nonumber.
\end{eqnarray}

R. Penrose proposed a Cosmic Censorship Conjecture (CCC) \cite{Pen1,Pen2,Pen3} to forbids the occurence of naked singularities in a solution of the field equations. According to the Weak Cosmic Censorship, singularities have no effect on distant observers, {\it i.e.}, they cannot communicate with far away observers. Till date, there is no mathematical theorems or proofs that support (or counter) this Conjecture. In contrary, there is no mathematical reason that a naked singularity cannot exist in a solution of the field equations. Therefore, the formulation and proof (or disproof) of this Conjecture remains one of the unsolved problems in General Relativity. Attempts have been made to provide the theoretical framework to devise a technique to distinguish between black holes and naked singularities from astrophysical data mainly through gravitational lensing (GL). Some significant works in this direction are the study of strong gravitational lensing in the Janis-Newman-Winicour spacetime \cite{KS,KS2} and its rotating generalization \cite{Yaz} and notably the work in \cite{Wer,Bambi,Bambi2,Hioki}. Other workers have shown that naked singularities and black holes  be differentiated by the properties of the accretion disks that accumulate around them. Consequently, the study of naked singularities and spacetime with such objects is of considerable current interest. In \cite{Chow}, the authors have enumerated three possible end states of gravitational collapse. There are number of examples of gravitational collapse in spherically symmetric that formed a naked singularities known. An earliest model that admits both naked and covered singularities is the Lemaitre-Tolman-Bondi (LTB) \cite{Lei,Lei1,Lei2} solutions, a spherically symmetric inhomogeneous collapse of dust fluid. Papapetrou \cite{pap} pointed out the formation of naked singularities in Vaidya \cite{vai} radiating solution, a null dust fluid spacetime generated from Schwarzschild vacuum solution. A small sample of spherically symmetric gravitational collapse solutions with a naked singularities are in \cite{Chri,Jos,Jos1,Jos2, Jos3,Jos4,Gov,Glass,Roc,Herr,Krasin}. Thus, the theoretical existence of naked singularities would mean that the gravitational collapse may observable from the rest of the spacetime.

In the present article, an axially symmetric solution of the field equations with a naked curvature singularity, will be presented. The Gravitational collapsing of cylindrically symmetric models that formed a naked singularity has been discussed in \cite{Thor,Hay}. Apostolatos and Thorne \cite{Apo} investigated the collapse of counter-rotating dust shell cylinder and showed that rotation, even if it is infinitesimally small, can halt the gravitational collapse of the cylinder. Echeveria \cite{Eche} studied the evolution of a cylindrical dust shell analytically at late times and numerically for all times. Guttia {\it et al} \cite{Gutt} studied the collapse of non-rotating, infinite dust cylinders. Nakao {\it et al} \cite{Nakao} studied the high-speed collapse of cylindrically symmetric thick shell composed of dust, and perfect fluid with non-vanishing pressure \cite{Nakao1}. Recent work describes the cylindrically symmetric collapse of counter-rotating dust shells \cite{Gonc,Nolan,Peri}, self-similar scalar field \cite{Wang}, axially symmetric vacuum spacetime \cite{Fa}, and a cylindrically symmetric anisotropic fluid spacetime \cite{Fa2}. Some other examples of non-spherical gravitational collapse would be discussed in \cite{Chi,Piran,Th,Morgan,Lete,JMM,Bondi,M,Wang2}.

The Einstein field equations (taking cosmological constant $\Lambda=0$) are given by
\begin{equation}
R_{\mu\nu}-\frac{1}{2}\,g_{\mu\nu}\,R=T_{\mu\nu},\quad \mu,\nu\in (0,1,2,3),
\label{1}
\end{equation}
where $R^{\mu\nu}$ is the Ricci tensor, $R$ is the Ricci scalar and $T^{\mu\nu}$ is the energy momentum tensor.

Pure radiation or Null dust fields are the fields of massless radiation which is considered as the incoherent superposition of waves with random phases and different polarizations but with the same propagation direction. The radiation can arise from fields of different types, from electro-magnetic null fields, massless scalar fields, neutrino fields or from the high frequency limit of gravitational waves. The energy-momentum tensor of pure radiation field \cite{Steph} is
\begin{equation}
T^{\mu\nu}=\rho\,k^{\mu}\,k^{\mu}=2\,\boldsymbol{\Phi_{22}}\,k^{\mu}\,k^{\nu},\quad k_{\mu}\,k^{\mu}=0,
\label{2}
\end{equation}
where $\rho$ is the energy density of null dust (pure radiation field) and $k^{\mu}$ is the null vector.

\section{Analysis of the Null dust Spacetime} 
  
Consider the following axially symmetric metric in $(t,r,\phi,z)$ coordinates
\begin{equation}
ds^2=g_{rr}\,dr^2+g_{\phi\phi}\,d\phi^2+2\,g_{t\phi}\,dt\,d\phi+2\,g_{z\phi}\,dz\,d\phi+g_{zz}\,dz^2,
\label{3}
\end{equation}
where the different metric functions are
\begin{eqnarray}
g_{rr}&=&\mbox{sech}^{4} r\,\mbox{tanh} r,\nonumber\\
g_{\phi\phi}&=&-\mbox{tanh}^{2} r\,\sinh t,\nonumber\\
g_{\phi t}=g_{t\phi}&=&g'_{\phi\phi},\nonumber\\
g_{\phi\,z}=g_{z\phi}&=&z\,\mbox{tanh}^{2} r,\nonumber\\
g_{zz}&=&\mbox{coth} r,
\label{4}
\end{eqnarray}
where prime deontes derivate w. r. to time, $g'_{\phi\phi}=\frac{dg_{\phi\phi}}{dt}$. Here $\phi$ coordinate is assumed periodic $\phi\in[0,2\,\pi)$. We have labelled the coordinates $x^0=t$, $x^1=r$, $x^2=\phi$ and $x^3=z$. The ranges of the other coordinates are $-\infty < t < \infty$, $0 \leq r < \infty$ and $-\infty < z < \infty$. The metric is Lorentzian with signature $(+,+,+,-)$ and the determinant of the corresponding metric tensor $g_{\mu\nu}$ is
\begin{equation}
det\;g=-\mbox{tanh}^{4} r\,\mbox{sech}^{4} r\,\cosh^{2} t.
\label{5}
\end{equation}
For spacetime (\ref{3}), the Ricci scalar $R=0$ and the non-zero component of the Ricci tensor $R^{\mu\nu}$ is 
\begin{equation}
R_{\phi\phi}=\frac{1}{2}\,\mbox{tanh}^{3} r.
\label{6}
\end{equation}

For spacetime (\ref{3}), the null vector is defined by $k_{\mu}=(0,0,1,0)$. Therefore, the non-zero component of the energy-momentum tensor (\ref{2}) is 
\begin{equation}
T_{\phi\phi}=\rho.
\label{8}
\end{equation}
From the field equations (\ref{1}) using (\ref{6}) and (\ref{8}), we get  
\begin{equation}
\rho(r)=\frac{1}{2}\,\mbox{tanh}^{3} r.
\label{9}
\end{equation}
The energy density of null dust satisfies the null energy condition (NEC).

The curvature scalar invariants for metric (\ref{3}) are
\begin{eqnarray}
R^{\mu\nu\rho\sigma}\,R_{\mu\nu\rho\sigma}&=&12\,\mbox{coth}^{6} r,\nonumber\\
R^{\mu\nu\rho\sigma;\tau}\,R_{\mu\nu\rho\sigma;\tau}&=&180\,\mbox{coth}^{9} r.
\label{7}
\end{eqnarray}
From the above equation, it is clear that the scalar curvature invariants and first-order differential invariants diverge (blow up) on the symmetry axis $r=0$, indicating the existence of a naked curvature singularity. Since the naked curvature singularity occurs without an event horizon, the Cosmic Censorship Conjecture has no physical interest.

\subsection{Geodesics in the Neighborhood of the Singularity}

To discuss the geodesics motion of free test particles which necessarily hit the singularity $r=0$, one needs to derive expression for their paths. Here we focus on radial geodesics on the symmetry axis $r=0$ \cite{Kurita}.

The Lagrangian for the metric (\ref{3}) is given by
\begin{eqnarray}
\nonumber
\pounds&=&\frac{1}{2}\,g_{\mu\nu}\,\dot{x}^{\mu}\,\dot{x}^{\nu}\\
&=& \frac{1}{2}\,\left[ g_{rr}\,\dot{r}^2+g_{zz}\,\dot{z}^2+g_{\phi\phi}\,\dot{\phi}^2+2\,g_{t\phi}\,\dot{t}\,\dot{\phi}+2\,g_{z\phi}\,\dot{z}\,\dot{\phi}\right ],
\label{15}
\end{eqnarray}
where dot stands derivative w. r. to an affine parameter. From (\ref{3}) and (\ref{15}), it is clear that $\phi$ is a cyclic coordinate. There exist constant of motion corresponding to this cyclic coordinate, {\it i.e.}, the azimuthal angular momentum $p_{\phi}$ is a constant given by 
\begin{equation}
p_{\phi}=const=g_{z\phi}\,\dot{z}+g_{\phi\phi}\,\dot{\phi}+g_{t\phi}\,\dot{t}.
\label{16}
\end{equation}

For the metric (\ref{3}), geodesics equation for $t$, $r$ coordinates are 
\begin{eqnarray}
\ddot{t}&=&\mbox{sech} t\,\{-\frac{1}{2}\,z^2\,\mbox{tanh}^{3} r\,\dot{\phi}^2+\dot{z}\,(3\,z\,\mbox{csch} r\,\mbox{sech} r\,\dot{r}+\dot{z})\}\nonumber\\
&-&(2\,\mbox{csch} r\,\mbox{sech} r\,\dot{r}+\dot{\phi})\,\dot{t}-\frac{1}{2}\,\mbox{tanh} t\,(\dot{\phi}^2+2\,\dot{t}^2),
\label{18}
\end{eqnarray}
\begin{eqnarray}
\ddot{r}&=&\frac{1}{2}\,[-\mbox{sech} r\,\mbox{csch} r\,\dot{r}^2+4\,\mbox{tanh} r\,\dot{r}^2\nonumber\\
&-&\cosh^{2}\,r\,\{2\,\sinh t\,\dot{\phi}^2-4\,z\,\dot{\phi}\,\dot{z}+\mbox{coth}^3 r\,\dot{z}^2+4\,\cosh t\,\dot{\phi}\,\dot{t}\}].
\label{19}
\end{eqnarray}

For radial geodesics we have $\dot{z}=0=\dot{\phi}$. From eqns. (\ref{18}), (\ref{19}) we get
\begin{eqnarray}
\ddot{t}&=&-2\,\mbox{csch} r\,\mbox{sech} r\,\dot{r}\,\dot{t}-\mbox{tanh} t\,\dot{t}^2,\nonumber\\
\ddot{r}&=&\frac{1}{2}\,(-\mbox{csch} r\,\mbox{sech} r+4\,\mbox{tanh} r)\,\dot{r}^2.
\label{21}
\end{eqnarray}
The solution of the above equations (\ref{21}) are
\begin{eqnarray}
\dot{t}(s)&=&c_1\,\mbox{sech} t\,\mbox{coth}^2 r,\quad c_1>0,\nonumber\\
\dot{r}(s)&=&c_2\,\cosh^{2} r\,\mbox{coth}^{1/2} r,\quad c_2>0.
\label{22}
\end{eqnarray}
Again solving the equations (\ref{22}) yields 
\begin{eqnarray}
t(s)&=&\sinh^{-1} [c_4-\frac{3\,c_1}{c_2}\,(2/3)^{4/3}\,(c_3+c_2\,s)^{-1/3}],\nonumber\\
r(s)&=&\mbox{tanh}^{-1} [(\frac{3}{2})^{2/3})\,(c_3+c_2\,s)^{2/3}],
\label{24}
\end{eqnarray}
where $c_i, i=1\ldots 4$ are arbitrary constants. Thus from the eqn. (\ref{24}), it is clear that the geodesic path $t$ is complete for finite value of the affine parameter $s$ except at $s=0$ (taking $c_3=0$), where it is unbounded. Therefore, the presented spacetime is radially geodesically incomplete.

\subsection{Strength of Naked Singularities}

The strength of singularity, which is the measure of its destructive capacity, is the most important feature. To determine the strength of naked singularities as strong and weak types, we consider the criteria developed by Tipler \cite{Tipler} and Krolak \cite{Kro}, which provide insights on the magnitude of tidal forces experienced by an in-falling detector (or an observer) towards the singularity \cite{Clarke2}. It is widely believed, that the analytical extension of the spacetime through a singularity is not possible, if it satisfies the strong curvature condition.

A naked singularity (NS) is said to be strong if the collapsing objects do get crushed to a zero volume at the singularity, and a weak one if they do not \cite{Jos1,Dwi}. Following Clarke and Krolak \cite{Clarke2}, a sufficient condition for a singularity to be strong in the sense of Tipler \cite{Tipler} is that at least along one radial geodesic we must have
\begin{equation}
\lim_{s\rightarrow 0} s^2\,R_{\mu\nu}\,\frac{dx^{\mu}}{ds}\,\frac{dx^{\nu}}{ds}\neq 0(>0),
\label{25}
\end{equation}
where $\frac{dx^{\mu}}{ds}$ is the tangent vector to the radial geodesics and $R_{\mu\nu}$ is the Ricci tensor. While
the weaker condition, which we called the {\it limiting focusing condition} \cite{Kro} is defined by
\begin{equation}
\lim_{s\rightarrow 0} s\,R_{\mu\nu}\,\frac{dx^{\mu}}{ds}\,\frac{dx^{\nu}}{ds}\neq 0.
\label{26}
\end{equation}

Hence from eqn. (\ref{25}), we have the strong curvature condition for the spacetime (\ref{3})
\begin{eqnarray}
&&\lim_{s\rightarrow 0} s^2\,[R_{rr}\,(\frac{dr}{ds})^2+R_{tt}\,(\frac{dt}{ds})^2]\nonumber\\
&=&\lim_{s\rightarrow 0} s^2\,[0\times (\frac{dr}{ds})^2+0\times (\frac{dt}{ds})^2]\nonumber\\
&=&0.\nonumber
\end{eqnarray}
Similarly, one can calculate the {\it limiting focusing condition}, and it also becomes zero. Thus the naked singularity (NS) which is formed due to the curvature singularity satisfies neither the strong curvature condition nor the {\it limiting focusing condition}. Therefore, the analytical extension of the spacetime through the singularity is possible.

\subsection{Equation of Orbit and Gravitational Lensing}

We calculate first the equation of orbit for $r(\phi)$. Here we restrict our discussion to the orbital motion of the free test particle which moves in the $z=const$-plane. To get an equation for $r(\phi)$ we start with
\begin{equation}
r'(\phi)=\frac{dr}{d\phi}=\frac{\dot{r}}{\dot{\phi}}.
\label{orbit}
\end{equation}
To work out the lens equation we have to calculate the null geodesics in $z=const$-planes. From the null geodesics $ds^2=0$ using the spacetime (\ref{3}) we have
\begin{equation}
g_{rr}\,\dot{r}^2+g_{\phi\phi}\,\dot{\phi}^2+2\,g_{t\phi}\,\dot{t}\,\dot{\phi}=0,
\label{27}
\end{equation}
where we have chosen $z=z_0$ (where $z_0$, a constant equal to zero).

We define angular velocity $\Omega$ of a zero angular momentum particle as measured by an observer (ZAMO) in $z=const$-planes. The angular velocity $\Omega$ (the angular velocity of the frame dragging) is define by \cite{Coll,Mei}
\begin{equation}
\Omega=\frac{\dot{\phi}}{\dot{t}}=-\frac{g_{t\phi}}{g_{\phi\phi}}=-\frac{g'_{\phi\phi}}{g_{\phi\phi}}=-\mbox{coth} t,\quad t<0.
\label{angular}
\end{equation}
Therefore, from eqn. (\ref{27}) using eqn. (\ref{angular}), one can obtain the following equation for the photon orbit
\begin{eqnarray}
&&g_{rr}\,\dot{r}^2-g_{\phi\phi}\,\dot{\phi}^2=0\nonumber\\\Rightarrow
&&(\frac{\dot{r}}{\dot{\phi}})^2=\frac{g_{\phi\phi}}{g_{rr}}\nonumber\\\Rightarrow
&&\frac{\dot{r}}{\dot{\phi}}=\pm\,\sqrt{\frac{g_{\phi\phi}}{g_{rr}}}\nonumber\\\Rightarrow
&&\frac{dr}{d\phi}=\pm\,\sqrt{\frac{g_{\phi\phi}}{g_{rr}}}=\pm\,\sqrt{H(r,t)}.
\label{28}
\end{eqnarray}
Substituting $g_{\phi\phi}$ and $g_{rr}$ and integration of eqn. (\ref{28}) immediately gives the azimuthal angle $\phi$ as a function of $r$
\begin{eqnarray}
\phi&=&\pm\,2\,\sqrt{-\mbox{csch} t}\,\mbox{tanh}^{1/2} r\nonumber\\
&=&\pm\,2\,\mbox{csch}^{1/2} T\,\mbox{tanh}^{1/2} r,
\label{29}
\end{eqnarray}
where $t=-T<0$, $T>0$. Here we have set $\phi(r=\infty)=0$. Thereofore, the Einstein deflection angle $\hat{\boldsymbol \alpha}$ is 
\begin{eqnarray}
\hat{\boldsymbol \alpha}(r_0)&=&2\,\int_{r=r_0}^{\infty}\,{\frac{dr}{\sqrt{H(r,t)}}}-\pi\nonumber\\
&=&4\,\mbox{csch}^{1/2} T\,(1-\mbox{tanh}^{1/2} r_0)-\pi.
\label{30}
\end{eqnarray}

\subsection{Closed Timelike Curves of the Spacetime: A Cosmic Time Machine}

The presented spacetime admits closed timelike curves which appear after a certain instant of time. There are many solutions of the field equations admitting closed timelike curves known (see references in \cite{Fa3}).

Consider an azimuthal curve $\gamma$ defined by $r= r_0,z=z_0,t=t_0$, where $r_0>0$, $z_0$ and $t_0$ are constants. From the line element (\ref{3}) we have 
\begin{equation}
ds^2=-\sinh t\,\mbox{tanh}^{2} r\,d\phi^2.
\label{31}
\end{equation}
These curves are null at $t=t_0=0$, spacelike throughout $t=t_0<0$, but become timelike for $t=t_0>0$, which indicates the presence of closed timelike curves (CTCs). Hence CTCs form at a definite instant of time satisfying $t=t_0>0$. The above analysis is valid provided that the CTCs evolve from an initial spacelike $t=const$ hypersurface (thus $t$ is a time coordinate) \cite{Ori1}. This can be determined by calculating the norm of the vector $\nabla_\mu t$ (or by determining the sign of $g^{tt}$ in the metric tensor $g^{\mu\nu}$ \cite{Ori1}). A hypersurface $t=const$ is spacelike provided $g^{tt}<0$ for $t<0$, timelike provided $g^{tt}>0$ for $t>0$ and null hypersurface $g^{tt}=0$ for $t=0$. In our case, from metric (\ref{3}) we found that
\begin{equation}
g^{tt}=\nabla_\mu t \nabla^\mu t=\mbox{sech}^{2} t\,\mbox{coth}^{2} r\,(z^2\,\mbox{tanh}^{3} r+\sinh t).
\label{32}
\end{equation}
Here we have chosen constant $z-planes$ defined by $z=z_0$, where $z_0$, a constant equal to zero. Thus, the hypersurface $t=const=t_0<0$ is spacelike ($r\neq 0$) and can be chosen as initial conditions over which the initial data may be specified. There is a Cauchy horizon at $t=t_0=0$ for any such initial spacelike hypersurface $t=t_0<0$. The null curve $t=t_0=0$ serves as the Chronolgy horizon which separates the spacetime a non-chronal region without CTCs to a chronal region with CTCs. The Chronology horizon is a special type of Cauchy horizon has been discussed detailed in \cite{Haw,Hawking}. Hence the spacetime evolves from an initial spacelike hypersurface in a causally well-behaved, up to a moment, {\it i.e.}, a null hypersurface $t=t_0=0$, and the formation of CTCs takes place from a causally well-behaved initial conditions.

The possibility that a naked curvature singularity gives rise to a Cosmic Time Machine has been discussed by Clarke and de Felice \cite{Clark} (see also,\cite{Felice,Felice2,Felice3}). A Cosmic Time Machine is a spacetime which is asymptotically flat and admits closed non-spacelike curves which extend to future infinity. The author and collaborators constructed such Cosmic Time Machines \cite{Fa,Fa2}, quite recently. The presented time-dependent spacetime may represent such a Cosmic Time Machine.

\section{Classification of the Spacetime and Effects on the Test Particles}

For classification of the metric (\ref{3}), one can construct a set of null tetrad vectors ${(\bf k, \bf l, \bf m,\bf {\bar m})}$. The metric tensor can be expressed as
\begin{equation}
g_{\mu \nu}=-k_{\mu}\,l_{\nu}-l_{\mu}\,k_{\nu}+m_{\mu}\,\bar{m}_{\nu}+\bar{m}_{\mu}\,m_{\nu},
\label{33}
\end{equation}
where the tetrad vectors are orthogonal except for $k_{\mu}\,l^{\mu}=-1$ and $m_{\mu}\,{\bar m}^{\mu}=1$. The non-zero Weyl scalars using the set of null tetrad are
\begin{eqnarray}
\Psi_2&=&C_{\mu\nu\rho\sigma}\,k^{\mu}\,m^{\nu}\,{\bar m}^{\rho}\,l^{\sigma}=\frac{1}{2\,\mbox{tanh}^{3} r},\nonumber\\
\Psi_4&=&C_{\mu\nu\rho\sigma}\,l^{\mu}\,{\bar m}^{\nu}\,l^{\rho}\,{\bar m}^{\sigma}=-\frac{1}{4}\,\mbox{tanh}^3 r-i\,\frac{3\,z}{4}\,\mbox{tanh} r,
\label{34}
\end{eqnarray}
while others are $\Psi_0=0=\Psi_1=\Psi_3$.

We set up an orthonormal frame ${\bf e}_{(a)}=\{{\bf e}_{(0)}, {\bf e}_{(1)}, {\bf e}_{(2)}, {\bf e}_{(3)}\}$, ${\bf e}_{(a)}\cdot{\bf e}_{(b)}\equiv {e_{(a)}}^{\mu}\,{e_{(b)}}^{\nu}\,g_{\mu\nu}=\eta_{(a)(b)}=\mbox{diag}(-1,+1,+1,+1)$ which consists of one timelike vector ${\bf e}_{(0)}$ and three spacelike unit vectors ${\bf e}_{(i)}$, $i=1,2,3$. Notation are such that small Latin indices are raised and lowered with $\eta^{(a)(b)}$, $\eta_{(a)(b)}$, and Greek indices are raised and lowered with $g^{\mu \nu}$, $g_{\mu \nu}$. These frame components can conveniently be expressed using the corresponding null tetrad vectors as: 
\begin{eqnarray}
{\bf e}_{(0)}&=&\frac{1}{\sqrt{2}}({\bf k}+{\bf l}),\quad {\bf e}_{(1)}=\frac{1}{\sqrt{2}}({\bf m}+\bf {\bar m}),\nonumber\\
{\bf e}_{(2)}&=&\frac{1}{\sqrt{2}}({\bf k}-{\bf l}),\quad {\bf e}_{(3)}=\frac{-i}{\sqrt{2}}({\bf m}-\bf {\bar m}).
\label{35}
\end{eqnarray}

In order to analyze the effects of the gravitational field of the above vacuum solution, it is natural to investigate the specific influence of various components of these fields on the relative motion of the free test particles. Such a local characterization of spacetimes, based on the equation of geodesic deviation frame, was described by Pirani \cite{Pirani1,Pirani2} and Szekeres \cite{Szek1,Szek2} ( see also \cite{Jiri1,Jiri2})
\begin{equation}
D^{2}\,Z^{\mu}/d\tau^2=-R^{\mu}_{\,\nu\rho\sigma}\,U^{\nu}\,Z^{\rho}\,U^{\sigma},
\label{36}
\end{equation}
where ${\bf U}\cdot{\bf U}=-1$, is the timelike four-velocity of the free test particle (observer), and $Z(\tau)$ is the displacement vector, where $\tau$ is the proper time. By projecting (\ref{36}) onto the orthonormal frame ${\bf e}_{(a)}$ given by eq. (\ref{35}), we get 
\begin{equation}
\ddot{Z}^{(i)}=-R^{(i)}_{\,(0)(j)(0)}\,Z^{(j)},
\label{37}
\end{equation}
where $i,j=1,2,3$, ${\bf e}_{(0)}=U$ and 
\begin{equation}
R^{(i)}_{\,(0)(j)(0)}=e^{\,\mu}_{(i)}\,U^{\nu}\,e^{\,\rho}_{(j)}\,U^{\sigma}\,R_{\mu\nu\rho\sigma}.
\label{38}
\end{equation}
The frame components of the displacement vector $Z^{(j)}\equiv e^{(i)}_{\mu}\,Z^{\mu}$ determines directly the distance between close test particles. Their physical relative accelerations are given by
\begin{equation}
\ddot{Z}^{(i)}=e^{(i)}_{\mu}\,(D^{2}Z^{\mu}/d\tau^2).
\label{39}
\end{equation}
Eq.(\ref{36}) also implies 
\begin{equation}
\ddot{Z}^{(0)}=-U_{\mu}\,(D^{2}\,Z^{\mu}/d\tau^2)=R_{\mu\nu\rho\sigma}\,U^{\mu}\,U^{\nu}\,Z^{\rho}\,U^{\sigma}=0,
\label{40}
\end{equation}
so that $Z^{(0)}=a_0\,\tau+b_0$, $a_0, b_0$ are constants. Setting $Z^{(0)}=0$ all test particles are synchronized by the proper time. From the standard definition of the Weyl tensor using the metric (\ref{3}), we get 
\begin{equation}
R_{(i)(0)(j)(0)}=C_{(i)(0)(j)(0)}+\frac{1}{2}\,\{\delta_{ij}\,T_{(0)(0)}-T_{(i)(j)}\}.
\label{41}
\end{equation}

For spacetime (\ref{3}), the only non-vanishing scalars is given by (\ref{34}) so that
\begin{eqnarray}
C_{(1)(0)(1)(0)}&=&\frac{1}{2}\,\Re{e\Psi_4}-\Re{e\Psi_2},\quad C_{(2)(0)(2)(0)}=-\frac{1}{2}\,\Re{e\Psi_4}-\Re{e\Psi_2},\nonumber\\
C_{(1)(0)(2)(0)}&=&-\frac{1}{2}\,\Im{m\Psi_4},\quad C_{(3)(0)(3)(0)}=2\,\Re{e\Psi_2},\nonumber\\
T_{(0)(0)}&=&T_{(2)(2)}\equiv \boldsymbol{\Phi_{22}}.
\label{42}
\end{eqnarray}
Therefore, the equation of geodesic deviation (\ref{25}) takes the form
\begin{eqnarray}
\ddot{Z}^{(1)}&=&-R^{(1)}_{\,(0)(j)(0)}\,Z^{(j)}=-[C_{(1)(0)(1)(0)}+\frac{1}{2}\,T_{(0)(0)}]\,Z^{(1)}-C_{(1)(0)(2)(0)}\,Z^{(2)}\nonumber\\
&=&-\frac{1}{2}\,\Re{e\Psi_4}\,Z^{(1)}+\frac{1}{2}\,\Im{m\Psi_4}\,Z^{(2)}+\Re{e\Psi_2}\,Z^{(1)}-\frac{1}{2}\,\boldsymbol{\Phi_{22}}\,Z^{(1)}\nonumber\\
&=&-\boldsymbol{A_{+}}\,Z^{(1)}+\boldsymbol{A_{\times}}\,Z^{(2)}+\boldsymbol{\it C}\,Z^{(1)}-\frac{1}{2}\,\boldsymbol{\Phi_{22}}\,Z^{(1)},\\
\ddot{Z}^{(2)}&=&-R^{(2)}_{\,(0)(j)(0)}\,Z^{(j)}=-C_{(2)(0)(2)(0)}\,Z^{(2)}\nonumber\\
&=&\frac{1}{2}\,\Re{e\Psi_4}\,Z^{(2)}+\Re{e\Psi_2}\,Z^{(2)}=\boldsymbol{A_{+}}\,Z^{(2)}+\boldsymbol{\it C}\,Z^{(2)},\\
\ddot{Z}^{(3)}&=&-R^{(3)}_{\,(0)(j)(0)}\,Z^{(j)}=-[C_{(3)(0)(3)(0)}+\frac{1}{2}\,T_{(0)(0)}]\,Z^{(3)}\nonumber\\
&=&-2\,\boldsymbol{\it C}\,Z^{(3)}-\frac{1}{2}\,\boldsymbol{\Phi_{22}}\,Z^{(3)},
\end{eqnarray}
where $\boldsymbol{A_{+}}=\frac{1}{2}\,\Re{e\Psi_4}$, $\boldsymbol{A_{\times}}=\frac{1}{2}\,\Im{m\Psi_4}$, $\boldsymbol{\it C}=\Re{e\Psi_2}$, and $\Re{e}$, $\Im{m}$ stands for real and complex respectively. The above equations are well suited for physical interpretation. Clearly, the relative motions of nearby test particles depends on : 

\vspace{0.1cm}

1. the local free gravitational field, and consisting of two components. There is a Coulomb-component with amplitude $\boldsymbol {\it C}$ , and transverse wave components with amplitudes $\boldsymbol{A_{+}}$ and $\boldsymbol{A_{\times}}$ of two polarization modes ``$+"$ and ``$\times"$ representing the effect of gravitational waves on the particles in the presented type II spacetime.

\vspace{0.1cm}

2. the terms describing the matter-content pure radiation field $(\boldsymbol{\Phi_{22}})$ which affects the motion only in the transverse plane spanned by the vectors ${\bf e}_{(1)}$, ${\bf e}_{(3)}$.

\section{Conclusions} 

We presented an axially symmetric time-dependent solution of the field equations which possesses a naked curvature singularity on the symmetry axis. The naked singularity occurs without an event horizon, therefore the Cosmic Censorship has no physical interest. The spacetime satisfies null dust as that for matter-energy content with positive energy density which is finite on the symmetry axis and obeys the null energy condition. The radial geodesics of the spacetime near to the singularity was discussed and found that these are incomplete. The analytical extension of the spacetime through the singularity is possible since the strength of the naked curvature singularity fails to satisfy the strong curvature condition. We obtained the equation for the photon orbit and subsequently the deflection angle $\hat{\boldsymbol \alpha}$ which is found time-dependent. The presented spacetime admits closed timelike curves which develop at some particular moment from an initial spacelike hypersurface in a causally well-behaved manner. The possibility that a naked curvature singularity gives rise to a Cosmic Time Machine has been discussed by Clarke and de Felice \cite{Clark}. The presented spacetime may represent such a Cosmic Time Machine. Finally, the physical interpretation of the presented solution, based on the study of the equation of the geodesic deviation was presented. It was demonstrated that, the solution represents an exact gravitational waves consisting of two components, transverse wave components with amplitudes $\boldsymbol{A_{+}}$ and $\boldsymbol{A_{\times}}$ of two polarization modes ``$+"$ and ``$\times"$, and Coulomb-like components of amplitude $\boldsymbol {\it C}$. There is a matter-energy null dust field interaction with the test particles.

The author declares that there are no competing interests regarding the publication of this paper.

\end{document}